\documentclass[conference]{IEEEtran}
\usepackage{hyperref}
\IEEEoverridecommandlockouts
\usepackage{algorithmicx}
\usepackage{cite}
\usepackage{cuted}
\usepackage{svg}
\usepackage{amsmath,amssymb,amsfonts}
\usepackage{verbatim}
\usepackage{graphicx}
\usepackage{booktabs}
\usepackage[compatibility=false]{caption}
\usepackage{subcaption}
\usepackage{multirow}
\usepackage{balance}
\usepackage{balance}
\usepackage{siunitx}
\usepackage{enumitem}
\usepackage{threeparttable} 
\usepackage{float} 
\usepackage{wrapfig} 
\usepackage{mathtools} 
\usepackage{makecell} 

\usepackage{algpseudocode}
\usepackage[ruled]{algorithm2e}
\usepackage{listings}
\usepackage{array}
\usepackage{tikz}
\usetikzlibrary{shapes, arrows.meta, positioning, fit, backgrounds}
\usepackage{fontawesome5}
\usepackage{lipsum}
\usepackage{tabularx}
\usepackage{etoolbox}
\usepackage{tabularray}
\usepackage{colortbl}
\definecolor{lightgray}{gray}{0.84}
\usepackage{mathtools} 
\def\BibTeX{{\rm B\kern-.05em{\sc i\kern-.025em b}\kern-.08em
    T\kern-.1667em\lower.7ex\hbox{E}\kern-.125emX}}

\usepackage{tikz}
\usetikzlibrary{fit}
\tikzset{%
  highlight/.style={rectangle,rounded corners,fill=red!15,draw,fill opacity=0.5,thick,inner sep=0pt}
}

\makeatletter
\def\thickhline{%
  \noalign{\ifnum0=`}\fi\hrule \@height \thickarrayrulewidth \futurelet
   \reserved@a\@xthickhline}
\def\@xthickhline{\ifx\reserved@a\thickhline
               \vskip\doublerulesep
               \vskip-\thickarrayrulewidth
             \fi
      \ifnum0=`{\fi}}
\makeatother

\newlength{\thickarrayrulewidth}
\definecolor{headercolor}{gray}{0.90}
\definecolor{rowcolor}{gray}{0.97}
\definecolor{rulecolor}{gray}{0.7}

\begin{document}

\title{GaborFlow: Hardware-Software Co-Optimization of Gabor Filter Accelerator Design}


\author{\IEEEauthorblockN{Priyanka Agarwal, Pruthvi Parate, Raghav Khurana,
  Madhav Rao  }
\IEEEauthorblockA{International Institute of Information Technology Bangalore\\
Email: \{priyanka.agarwal, pruthvi.parate, raghav.khurana, mr\}@iiitb.ac.in}
}

\maketitle

\begin{abstract}
Brain tumor classification from MRI scans represents a critical challenge in medical imaging, where accuracy directly impacts patient outcomes. This paper introduces GaborFlow, a novel hardware-software co-optimization framework that revolutionizes the design of the Gabor Filter accelerator for generating feature maps to enhance brain tumor detection. Unlike conventional approaches that rely on fixed filter parameters, our method employs trainable Gabor coefficients within a carefully optimized range, integrated with an EfficientNet-B0 backbone CNN architecture. The innovation lies in the synergistic combination of software optimization through Bayesian hyperparameter tuning and hardware efficiency through precision-optimized multipliers. Our GF baseline accelerator achieves remarkable results, with 98.6\%  classification accuracy, 28.08\% power benefits, and 48.83\% footprint savings compared to state-of-the-art implementations, demonstrating that medical AI can be both highly accurate and energy-efficient. The hardware design files are made freely available for easy adoption and further usage by the researchers and designers.

\end{abstract}



\begin{IEEEkeywords}
Convolutional Neural Network (CNN), Non-dominated Sorting Genetic Algorithm-II (NSGA-II), VLSI architecture, Gabor Filter bank, Accelerator, Fixed point precision scaling
\end{IEEEkeywords}


\section{Introduction} \label{sec:intro}

Medical imaging, particularly Magnetic Resonance Imaging (MRI), has become a critical tool for diagnosing various brain disorders due to its high resolution and non-invasive nature~\cite{mri3, mri4, mri6}. Brain tumors such as glioma, meningioma, and pituitary tumors require accurate and early detection to improve patient outcomes~\cite{bt1, bt2, bt3, bt4}. Despite significant strides in medical imaging techniques, the automatic and precise classification of these tumors remains a hurdle due to variations in texture, intensity, and complex morphological characteristics. 


Convolutional Neural Networks (CNNs) have demonstrated remarkable success in medical imaging tasks, particularly in the classification and segmentation of brain MRI scans. Recent studies have highlighted the importance of integrating biologically inspired filters, like Gabor filters (GFs), into CNN architectures to enhance the networks' ability to capture texture and structural details efficiently. For instance, Naive Gabor Networks have shown significant parameter reduction and improved generalization by leveraging optimal time-frequency localization properties of the Gabor Filter (GF) in hyperspectral and general imaging tasks~\cite{naive-gabor}. Furthermore, Gabor Convolutional Networks demonstrate improved robustness to orientation and scale changes while drastically reducing network complexity~\cite{gabor, CNN1, CNN2}. In other applications, these biologically inspired networks have found particular success in challenging tasks, including pose-invariant face recognition, by enhancing discriminative features robustly across variations ~\cite{gabor-pose-invariant}. 

Recent studies have explicitly demonstrated improved segmentation accuracy by employing GFs, highlighting their potential to significantly enhance CNN-based brain tumor diagnostics~\cite{gabor-brain-mri, luan}. However, despite the promising performance of Gabor-augmented CNNs, existing literature primarily employs static, manually selected hyperparameters for the GFs. 
Moreover, there is scant research on integrating the hardware-optimized GF design to accelerate the CNN process for real-world deployment.
To address these gaps, this research proposes a fully trainable GF as the initial layer of a CNN model specifically optimized for brain tumor classification using MRI scans. We utilize the EfficientNet-B0 architecture, renowned for its exceptional balance between computational efficiency and classification performance. Previous works employing EfficientNet in medical imaging have consistently demonstrated remarkable classification accuracy and robustness in detecting brain tumors~\cite{efficientnetb0-main,efficientnetb0-transfer}. Bayesian optimization methods, specifically Optuna, are utilized to systematically explore the hyperparameter space and optimize the Gabor layer parameters based on a dual-objective criterion including F1-score and AUC-ROC, ensuring the model’s diagnostic reliability and clinical relevance~\cite{bayesian}. 

The coefficients of Gabor-wavelet filter banks (GWFB) are inherently irrational~\cite{gabor1}; hence, mapping them directly to hardware inflates memory usage, lowers throughput, and consumes more dynamic power. To mitigate these penalties, the coefficients are quantized to a finite-precision format while preserving accuracy. Prior work illustrates this trade-off: Cesur et al. \cite{cesur} expressed GF as separate real and imaginary FIR filters, using 12-bit fixed-point coefficients (FI-12), which supports video streams of up to 373.2 MP pixels/s. Licciardo et al. \cite{Licciardo} and Painkras \cite{Painkras} adopted 16-bit (FI-16) and 12-bit (FI-12) fixed-point formats, respectively; the latter architecture runs at 100 MHz, delivering a full Gabor filter bank~(GFB) in 2~ms per facial feature point, enabling 70 fps real-time face tracking. 
By contrast, Lee et al. \cite{Lee} used floating-point arithmetic, which demands power-hungry multipliers and Digital Signal Processing (DSP) slices. Aswini \cite{power-eff} and \cite{GF_int8} developed a hardware-efficient multiplierless GFB architecture based on 8-bit dyadic coefficients; however, the architecture in \cite{power-eff} is rigid and fixed for a specific application, and \cite{GF_int8} showcases a lower frame rate. Selecting an optimal numeric representation is, therefore, essential for hardware-efficient Gabor filtering. In our proposed work, we apply the Non-Dominated Sorting Genetic Algorithm II (NSGA-II) to tune the fixed-point precision of CNN-augmented Gabor kernels, resulting in a graded-precision coefficient set, tailored to each multiplier in the filter bank.
Contribution of the proposed work:

1) An EfficientNetB0$-$based CNN learning framework is employed to explore the hyperparameter space and optimize the GF parameters for the target dataset, thereby enhancing feature representation and improving the overall classification accuracy towards tumor detection~(4 classes).
This forms the Software optimization for GF Design.

2) A multi-objective meta-heuristic optimization method (NSGA$-$II) was used to find the hardware-efficient 
Graded precision GF representation (precision bit-widths unique to each multiplier), 
thereby enabling a more lightweight and efficient GF based feature extraction layer towards on-chip diagnosis. This forms the hardware optimization for the proposed design.

The extracted GF output is also validated with the EfficientNetB0 classification model. The workflow diagram is presented in the Figure~\ref{fig:modelarch}.
Overall, we propose Hardware-Software Co-optimization to 
deduce a hardware-efficient GF accelerator design. Beyond its application in medical imaging, our architecture demonstrates broad versatility and are effectively utilized in texture analysis, feature extraction, and edge detection.
The hardware design files are made freely available at~\cite{gaborsite_gaborflow} for further usage by the researchers and designers.


\section{Proposed Methodologies } \label{sec:prelims}

\subsection{GF Parameter Optimization}

\subsubsection{Gabor Filter Parameter Optimization}
GFs act as powerful feature extractors capable of capturing spatial and frequency-domain characteristics.
The 2D GF is mathematically expressed in the 
Equation~\ref{eq:gabor}, where the coordinate transformation accounts for orientation,
$x'= x\cos\theta + y\sin\theta$, 
$y'= -x\sin\theta+y\cos\theta$,

\begin{equation}
\begin{aligned}
g(x,y;\lambda,\theta,\phi,\sigma,\gamma)
&=\exp\!\left(
-\frac{x'^2+\gamma^2 y'^2}{2\sigma^2}
\right) \\
&\quad \times
\exp\!\left(
i\left(2\pi\frac{x'}{\lambda}+\phi\right)
\right)
\end{aligned}
\vspace{2mm}
\label{eq:gabor}
\end{equation}

The critical parameters that impact the functionality of  the filter are:
i) Wavelength ($\lambda$): Controls the scale of detected features, 
ii) Orientation ($\theta$): Referred to as directional sensitivity. Our method dynamically determines the optimal count,
iii) Standard deviation ($\sigma$): Gaussian envelope width,
iv) Aspect ratio ($\gamma$): Ellipticity of the Gaussian envelope, and
v) Phase offset ($\phi$): Shifts the sinusoidal component for enhanced feature detection. 

\begin{figure*}[!htp]
\centering
\vspace{-4mm}
\includegraphics[width=1\textwidth]{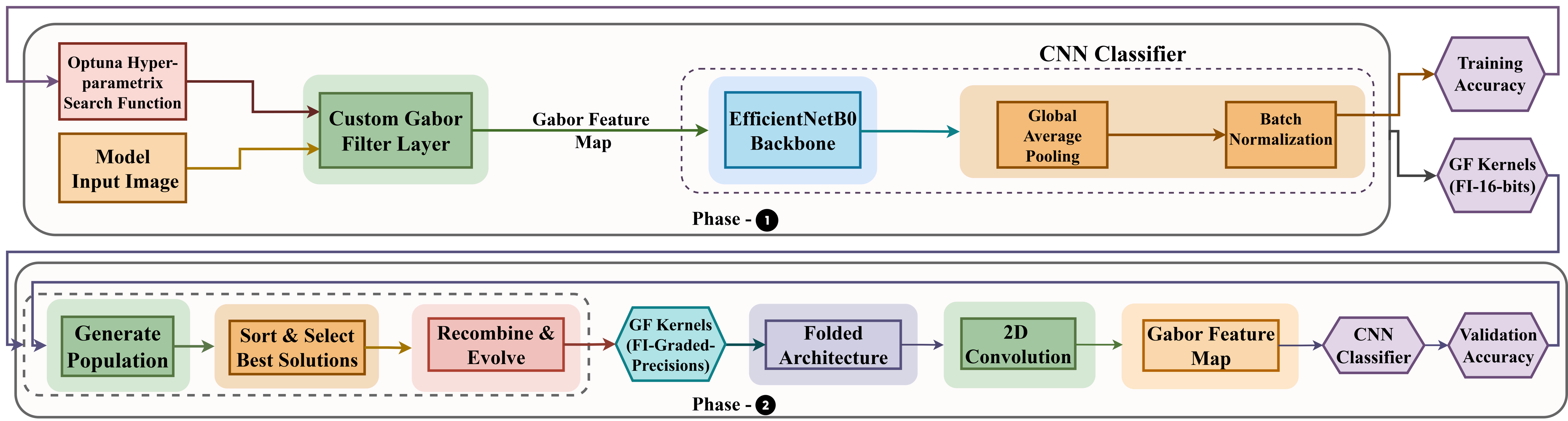}
 \vspace{1mm}
\caption{Proposed EfficientNet-Bo and NSGA-III based GF optimization methodology for feature map extraction.}
\label{fig:modelarch}
\vspace{-3mm}
\end{figure*}

The trainable nature of these parameters distinguishes the proposed approach from conventional fixed-parameter implementations, enabling dataset-specific optimization for brain tumor classification. The effectiveness of GF depends on the proper estimation of these parameters. We therefore propose a method to identify optimal configurations of these parameters based on a CNN model with a fully trainable GF as the initial layer, which is explicitly optimized for brain tumor classification using the Kaggle MRI scan dataset~\cite{kaggle_official}. 
Figure~\ref{fig:modelarch} shows our GaborFlow methodology that involves a two-phase workflow:
\begin{itemize}[leftmargin=*]
    \item \textit{Phase 1}: 
    The first stage involves optimizing GF coefficients by parameter extraction using EfficientNetB0-based training on the selected dataset.

    \item \textit{Phase 2}: Once the Gabor coefficients are established, the NSGA-II algorithm is used to determine the graded-precision multipliers tailored for the folded GF-accelerator architecture. These optimized multipliers are then used to generate features from the hardware-efficient GF, which are subsequently passed to the CNN for classification. This forms a more lightweight and efficient Filter layer for on-chip diagnosis. Graded Precision multipliers constitute the hardware optimization aspect in the proposed design.
\end{itemize}
For GF parameter extraction, we employed Bayesian hyperparameter optimization using the Optuna framework with k-fold cross-validation (4 folds) to maximize the macro F1-score. The GFB was initially configured with four $5\times 5$ kernels (basic light-weight model), with orientations sampled from narrow angular intervals to refine directional sensitivity. 
During optimization, the model was allowed to select the number of feature maps (orientations) required for best performance. The optimal configuration converged to 13 orientations.
Hence, we investigated both the light-weight four orientations (4OR) and optimal thirteen orientations (13OR) models as the proposed design.
The optimization space for the filter parameters includes the following:

\begin{itemize}[leftmargin=*]
    \item \textit{Sigma} $(\sigma)$: 3.363 to 3.383, with a step of 0.001
    \item \textit{Aspect Ratio} $(\gamma)$: 0 to 1, with a step of 0.01
    \item \textit{Wavelength} $(\lambda)$: 7.5 to 8.5, with a step of 0.05
    \item \textit{Orientations} $(\theta)$: Four and thirteen corresponding to base and optimum architectures.
\end{itemize}

These hyper-parameters were optimized over 10 trials to maximize classification performance on the validation set. Unlike traditional hand-crafted GFB, the implemented filter bank is \textit{fully differentiable} and \textit{trainable}, enabling end-to-end optimization during back-propagation.

\subsubsection{CNN Architecture with EfficientNet-B0 Backbone}
The proposed Gabor augmented CNN architecture, shown in Figure~\ref{fig:modelarch}, integrates a trainable GFB with EfficientNet-B0 model, with GF as its initial layer. 
We selected EfficientNet-B0 for its high accuracy-to-complexity ratio, making it suitable for potential hardware deployment. The CNN architecture processes $512\times512\times3$ MRI images using a trainable $5\times5 $ GFB with four orientations to enhance edge and texture features. A subsequent $1\times1$ convolutional layer is applied to ensure feature map compatibility with the EfficientNet-Bo backbone. An EfficientNet-B0 backbone pretrained on ImageNet is used, with the first 190 layers frozen and the final 50 layers fine-tuned to adapt to medical imaging requirements. 
The extracted features are processed through the Global Average Pooling and Batch Normalization layers to reduce the spatial dimensionality and stabilise the learning process. The data then passes through a 128-unit Dense layer, ReLU activation, and L2 regularization (coefficient 0.01), to learn complex feature interactions, followed by another Batch Normalization layer and a Dropout layer (rate 0.5) to prevent overfitting, culminating in a 4-class Softmax classification. 
The model is optimized with the Adam algorithm, using a learning rate derived from Bayesian optimization (divided by 10, ranging from $10^{-6}$ to $10^{-3}$) and gradient clipping (clipnorm=1.0) to ensure stable training. Categorical Cross-entropy is used as the loss function, which is suitable for multi-class classification.

The model was trained for 40 epochs per fold, with additional training on the full dataset using callbacks such as ModelCheckpoint (monitoring validation accuracy), EarlyStopping (patience of 10 epochs, monitoring validation loss), and ReduceLROnPlateau (factor of 0.5, patience of 5 epochs, minimum learning rate of $10^{-6})$. Class weights were applied to address imbalanced data.

\subsubsection{Classification Metrics and Model Evaluation}

The classifier's evaluation was conducted on a test dataset comprising 1755 samples across four distinct classes: glioma, meningioma, no tumor, and pituitary tumor~\cite{kaggle_official}. 
The GF design with 4 Orientations (GF-4OR) demonstrated robust ability to predict most tumor types, with high classification accuracy for no tumor (98\%) and pituitary tumor (100\%) classes. However, a lower sensitivity for meningioma (76\%) and glioma (89\%) indicated a need for additional data augmentation with higher feature maps. The classifier was trained to generate the optimum number of feature maps (13 in the proposed design) to enhance overall accuracy. The GF architecture with 13 orientations (GF-13OR) demonstrated high classification accuracy for all four classes.  Table~\ref{tab:classification-report} shows the normalized confusion matrix for the classifier, customized with a GF layer with 4 and 13 feature maps.

The customized EfficientNet-B0 classifier architecture was compared with state-of-the-art (SOTA) models in Table~\ref{tab:modelcomp}. The performance evaluation was conducted using GPU T4x2. VGG16 achieved 90\% accuracy with 14.7 million parameters and 15.71 ms inference time. ResNet50 showed improved accuracy with faster inference, but with 23.6 million parameters. Lightweight models, such as DenseNet121 and MobileNetV2, which had comparatively faster inference times, suffered a drop in accuracy. EfficientNet-B0 proved to be the best model choice, achieving an accuracy of 93\%, with only 4 million parameters and a fast inference time of 5.56 ms. Incorporating the GF-13OR into EfficientNet-B0 yielded an impressive accuracy of 98.6\% without any overhead on parameters and inference time.
In summary, the GaborFlow approach effectively combines GF with CNNs, demonstrating significant potential for enhancing brain tumor classification accuracy through hardware-software co-optimization.

Table~\ref{tab:dataset} reports the MRI image classification metrics for the GF-integrated CNN model on three different medical datasets~\cite{kaggle_official, kidney, lung}. The results indicate that integrating the GF layer into the CNN consistently outperforms the baseline EfficientNet-B0 across all imaging modalities, resulting in a 5.6\% increase in accuracy for brain tumor detection. This improvement highlights the GF layer’s capacity to extract enhanced edge and texture features, which are essential for distinguishing complex patterns in medical images.

\begin{table}[t]
\centering
\caption{ MRI Image classification report for the customized EfficientNet-B0 model. Here, 4OR represents four filters with four distinct orientations, while 13OR represents thirteen filters with distinct orientations.}
\label{tab:classification-report}

\resizebox{\columnwidth}{!}{%
\begin{tabular}{lcccccccc}
\toprule
\multirow{2}{*}{{Image Class}} 
& \multicolumn{2}{c}{{Specificity}} 
& \multicolumn{2}{c}{{Sensitivity}} 
& \multicolumn{2}{c}{{F1-Score}} 
& \multicolumn{2}{c}{{Support}} \\
\cmidrule(lr){2-3}
\cmidrule(lr){4-5}
\cmidrule(lr){6-7}
\cmidrule(lr){8-9}
& {4OR} & {13OR}
& {4OR} & {13OR}
& {4OR} & {13OR}
& {4OR} & {13OR} \\
\midrule

Glioma     
& 0.96 & \textbf{1.00}
& 0.89 & \textbf{0.97}
& 0.92 & \textbf{0.94}
& 549  & \textbf{550} \\

Meningioma 
& 0.92 & \textbf{0.98}
& 0.76 & \textbf{0.99}
& 0.84 & \textbf{0.89}
& 409  & 409 \\

No tumor   
& 0.90 & \textbf{0.99}
& 0.98 & \textbf{0.99}
& 0.94 & \textbf{0.95}
& 319  & 319 \\

Pituitary  
& 0.86 & \textbf{0.99}
& \textbf{1.00} & 0.99
& 0.92 & \textbf{0.98}
& 478  & 478 \\

\bottomrule
\end{tabular}%
}

\vspace{1mm}
\begin{minipage}{\columnwidth}
\footnotesize
\textit{Note:} Bold values indicate the best value for each metric.
\end{minipage}

\vspace{1mm}
\end{table}


\begin{table}[t]
\centering
\caption{Comparison of model performance metrics with state-of-the-art models.}
\label{tab:modelcomp}
\vspace{1mm}

\resizebox{\columnwidth}{!}{%
\begin{tabular}{lccc}
\toprule
  {Classifier Model} 
& {Accuracy (\%)} 
& {Parameters (M)} 
& {Inference Time (ms)} \\
\midrule

VGG16                     & 90.00 & 14.7 & 15.71 \\
ResNet50                  & 95.00 & 23.6 & 8.80  \\
DenseNet121               & 86.00 & 7.0  & 8.50  \\
MobileNetV2               & 87.00 & 2.2  & 4.52  \\
EfficientNet-B0           & 93.00 & 4.0  & 5.56  \\
EfficientNet-B0 (GF-4OR)  & 94.75 & 4.0  & 5.60  \\

\textbf{EfficientNet-B0 (GF-13OR)} 
& \textbf{98.60} & \textbf{4.0} & \textbf{5.60} \\
\bottomrule
\end{tabular}%
}

\vspace{-2mm}
\end{table}




\begin{table}[t]
\centering
\caption{Performance comparison of the proposed model with EfficientNet-B0 baseline on different medical datasets.}
\label{tab:dataset}
\vspace{1mm}

\resizebox{\columnwidth}{!}{%
\begin{tabular}{llcccc}
\toprule
{Model} & {Dataset} & {Accuracy (\%)} 
& {Sensitivity} & {Specificity} & {AUC} \\
\midrule

EfficientNet-B0 (Baseline) & \cite{kaggle_official} & 93.00 & 0.9805 & 0.9939 & 0.9986 \\
\textbf{EfficientNet-B0 (GF-13OR)} & \cite{kaggle_official} & \textbf{98.60} & \textbf{0.9958} & \textbf{0.9988} & \textbf{0.9995} \\
\midrule

EfficientNet-B0 (Baseline) & \cite{lung} & 90.77 & 0.9078 & 0.9692 & 0.9900 \\
\textbf{EfficientNet-B0 (GF-13OR)} & \cite{lung} & \textbf{94.28} & \textbf{0.9428} & \textbf{0.9809} & \textbf{0.9961} \\
\midrule

EfficientNet-B0 (Baseline) & \cite{kidney} & 98.47 & 0.9847 & 0.9847 & 0.9987 \\
\textbf{EfficientNet-B0 (GF-13OR)} & \cite{kidney} & \textbf{99.27} & \textbf{0.9927} & \textbf{0.9927} & \textbf{0.9995} \\

\bottomrule
\end{tabular}%
}

 \vspace{-1mm}
\end{table}







\begin{figure*}[!htp]
\vspace{-3.5mm}
\centering
\includegraphics[scale=0.060]{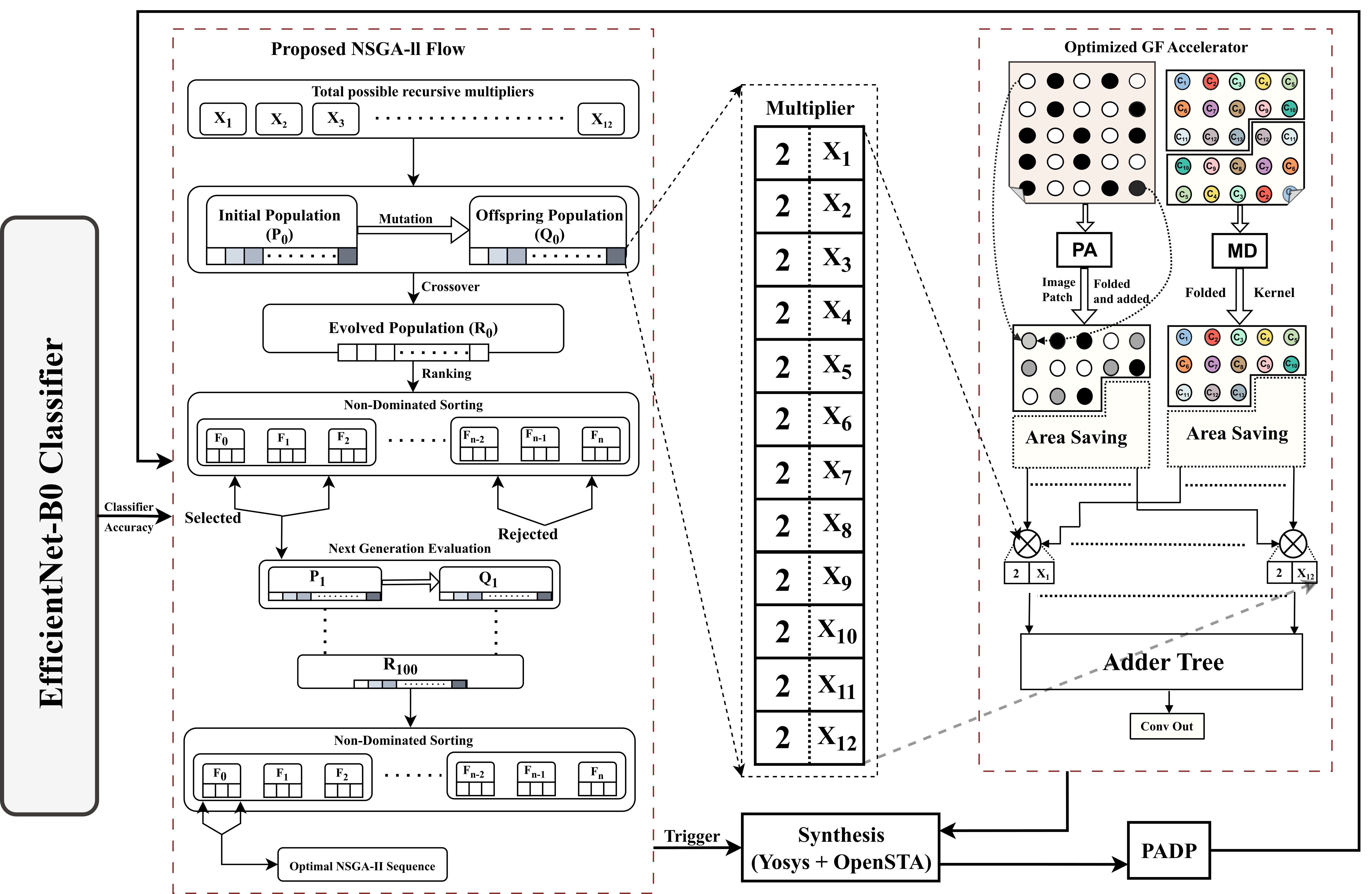}
\vspace{2mm}
\caption{Schematic Representation 
of the NSGA-II engineered Gabor Filter Architecture.}
\label{fig:gabornsga}
\end{figure*}

\subsection{Multiplier Optimization using NSGA-II Framework}
Phase 2 of our GaborFlow methodology, shown in Figure \ref{fig:modelarch}, suggests hardware optimization of the GF feature extraction layer using the NSGA-II algorithm. The GF in our implementation employs a Folded Kernel-based architecture due to coefficient symmetry along a given kernel. The folded technique was adopted to reduce the coefficient storage memory by $50\%$ and the number of multipliers to half for a given orientation. The filter design described in the next section, developed for tumor detection in MRI images, utilizes a GF with 5$\times$5 kernels. This architecture is generalized for any given orientation. Due to folding, the effective number of unique coefficients to be approximated in the 5$\times$5 kernel reduces to 13, where the thirteenth coefficient value is '1' for all filters.

To optimize the 12 multipliers in each kernel, we applied a customized NSGA-II algorithm~\cite{nsga_srin, NSGA2} that balances silicon area, delay, power, and classifier validation accuracy. Before running the evolutionary search, we pruned the design space by testing reduced bit-width multipliers. This analysis demonstrated that a 9-bit baseline (down from 16 bits) achieves comparable image quality (with an average SSIM of 0.94) while significantly reducing hardware costs, making it a practical starting point for further optimization.


The NSGA-II optimization algorithm was implemented to find the optimal set of graded fixed-point precisions for all $12$ multipliers, with the search space ranging from precisions of $3$ to $9$, considering a sign bit, $1$ integer bit, and 
fractional bits ranging from 1 to 7.
Each individual in the population represents a specific sequence of multiplier bit-widths for $12$ stages as shown in the Figure~\ref{fig:gabornsga}. 
The proposed NSGA-II flow begins with an initial population $P_{0}$ composed of candidate design solutions, where each encodes a sequence of multiplier widths across 12 stages. This population undergoes genetic operations, such as crossover and mutation, which are explained below, to produce an offspring population $Q_{0}$.

A) \textit{Crossover and Mutation:} Each candidate solution is encoded as a 12-dimensional integer vector, where each gene (multiplier) takes a value in $[3,9]$, representing its bit-width. This yields a vast design space of $7^{12}$ ($1.38 \times 10^{10}$) unique configurations, ensuring architectural diversity. New offsprings are generated through uniform crossover, where tournament-selected parents exchange genes. Point mutation introduces random perturbations within the range of $[3,9]$ to maintain diversity and prevent local optima, with a decaying mutation rate that enables broad early exploration.

Both populations are combined into $R_{0}$ and ranked using non-dominated sorting, which divides solutions into fronts $F_{0}, F_{1}, F_{2},...$ based on Pareto dominance. The first front $F_{0}$ contains the best non-dominated solutions, while later fronts include dominated ones. The best individuals from these fronts form the next generation $P_{1}$, which again generates offspring $Q_{1}$.
This cycle of combining, sorting, selecting, and reproducing continues over generations, gradually evolving the population towards optimal solutions. The final output is the NSGA-II sequence, which represents the Pareto front solution for the multi-objective optimization problem.


 \section{Hardware Implementation}\label{sec:proposed-method}

The proposed GF accelerator architecture employs an optimized FI-graded precision scheme, where each multiplier is assigned a precision level tailored to its corresponding coefficient. Figure~\ref{fig:gabornsga} illustrates the convolution of a $5\times5$ Gabor kernel with an input image and the corresponding hardware structure. The convolution module is implemented in Verilog with parameterizable FI precision, allowing flexible control of coefficient bit widths.
A key architectural insight comes from exploiting structural symmetries within the Gabor filter. For any orientation, diagonally opposite kernel coefficients are identical, and the central coefficient consistently takes the value ‘1’. Using this property, the \textit{Matrix Decomposer (MD)} identifies and extracts the unique coefficients and arranges them into a compact 
$1\times13$ vector representation, where the thirteenth element corresponds to the fixed center coefficient. This transformation effectively “folds” the original $5\times5$ kernel into a reduced form that preserves functional equivalence while minimizing redundancy.

To maintain alignment with the folded coefficient structure, the \textit{Pixel Accumulator (PA)} performs a complementary operation on the input image patch. Pixel pairs associated with symmetric kernel elements, such as $(\mathbf{P_1}$ and $\mathbf{P_{25}}$, are summed to form a single representative value. The PA thus converts the input $5\times5$ pixel neighborhood into a corresponding $1\times13$ folded pixel vector. By performing these pixel summations upfront, the architecture avoids redundant multiplications. For example, instead of individually computing $(\mathbf{P_1} \cdot \mathbf{C_1})$ + $(\mathbf{P_{25}}\cdot \mathbf{C_{25}})$, the PA enables a single simplified computation $(\mathbf{P_1} + \mathbf{P_{25}})\cdot \mathbf{C_1}$.
This folding strategy reduces coefficient memory by 50\% and cuts the number of required multipliers by 52\%, without impacting the kernel’s representational capability. The folded design also aligns neatly with the FI-graded precision framework.
Multiplication within the convolution block is performed using exact signed Dadda multipliers~\cite{bhargavanna}. The bit widths of these multipliers are assigned based on the precision values generated by the NSGA-II optimization algorithm, which selects the most efficient precision configuration for each coefficient while maintaining acceptable classification accuracy. Together, these optimizations yield significant improvements in chip area, power consumption, and computational throughput, enabling an efficient yet accurate hardware implementation of Gabor filtering.

\section{Experimental Setup and Results}\label{sec:results}


\begin{table}[t]
\centering
\caption{Hardware parameters used in the NSGA-II experimental setup.}
\label{tab:hardware_parameters}

\begin{tabular}{ll}
\toprule
{Parameter} & {Value} \\
\midrule

Technology Node &
Nangate 45\,nm (Typical Corner) \\

Clock Frequency &
100\,MHz, 50\% duty cycle \\

Input Activity Factor &
0.25 \\

External Input/Output Delay &
0\,ns \\

Input Driving Cell &
Buffer Cell Drive $1\times$ \\

Output Load &
6.80092\,fF (FO4 inverter) \\

\bottomrule
\end{tabular}

\vspace{1mm}
\end{table}

 



\begin{figure}[t]
    \centering
    \begin{minipage}[a]{0.42\textwidth}
        \includegraphics[width=7.5cm, height=4.5cm]{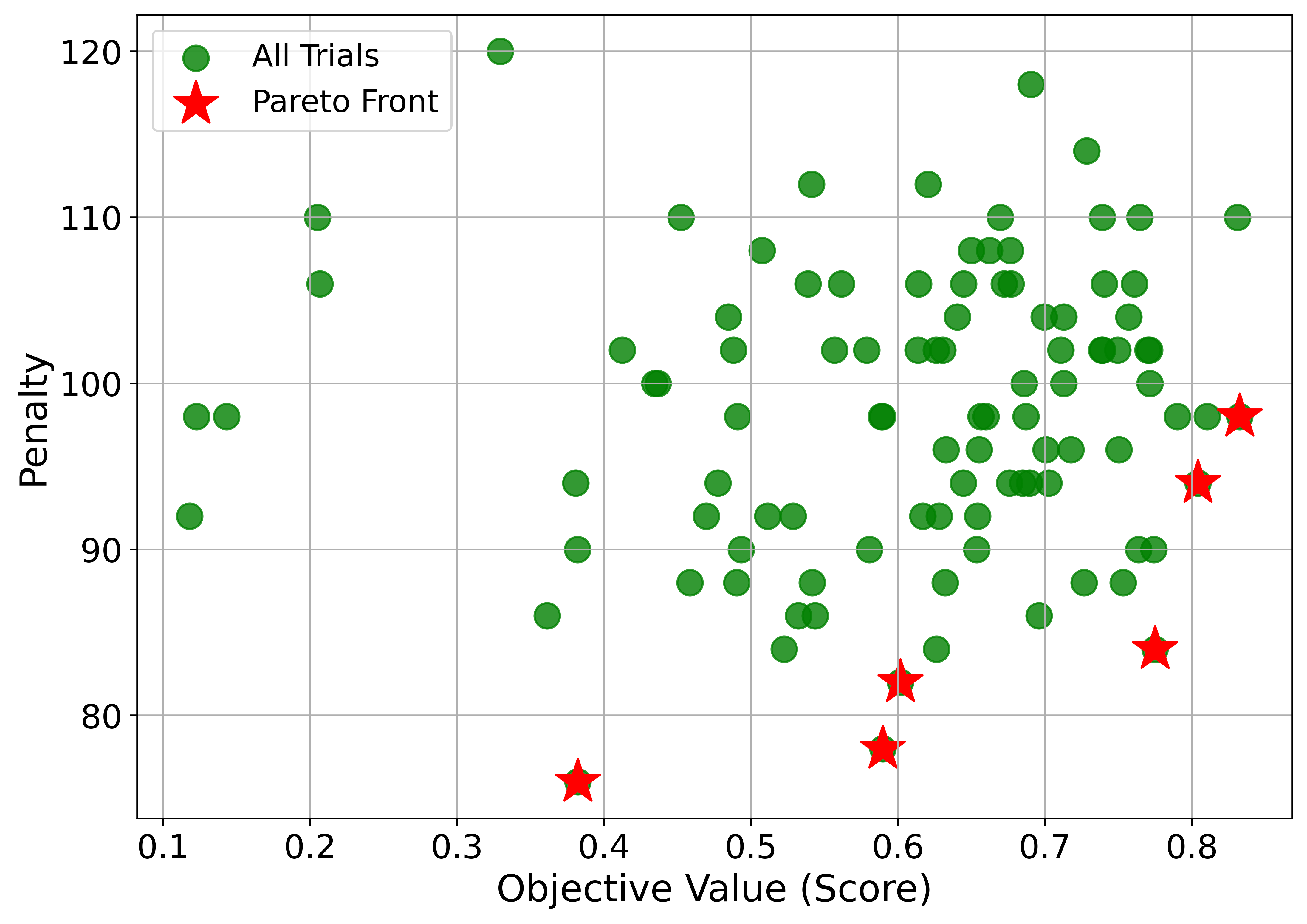}
        \subcaption{}
    \end{minipage}
    
    \begin{minipage}[b]{0.42\textwidth}
        \includegraphics[width=7.5cm, height=4.5cm]{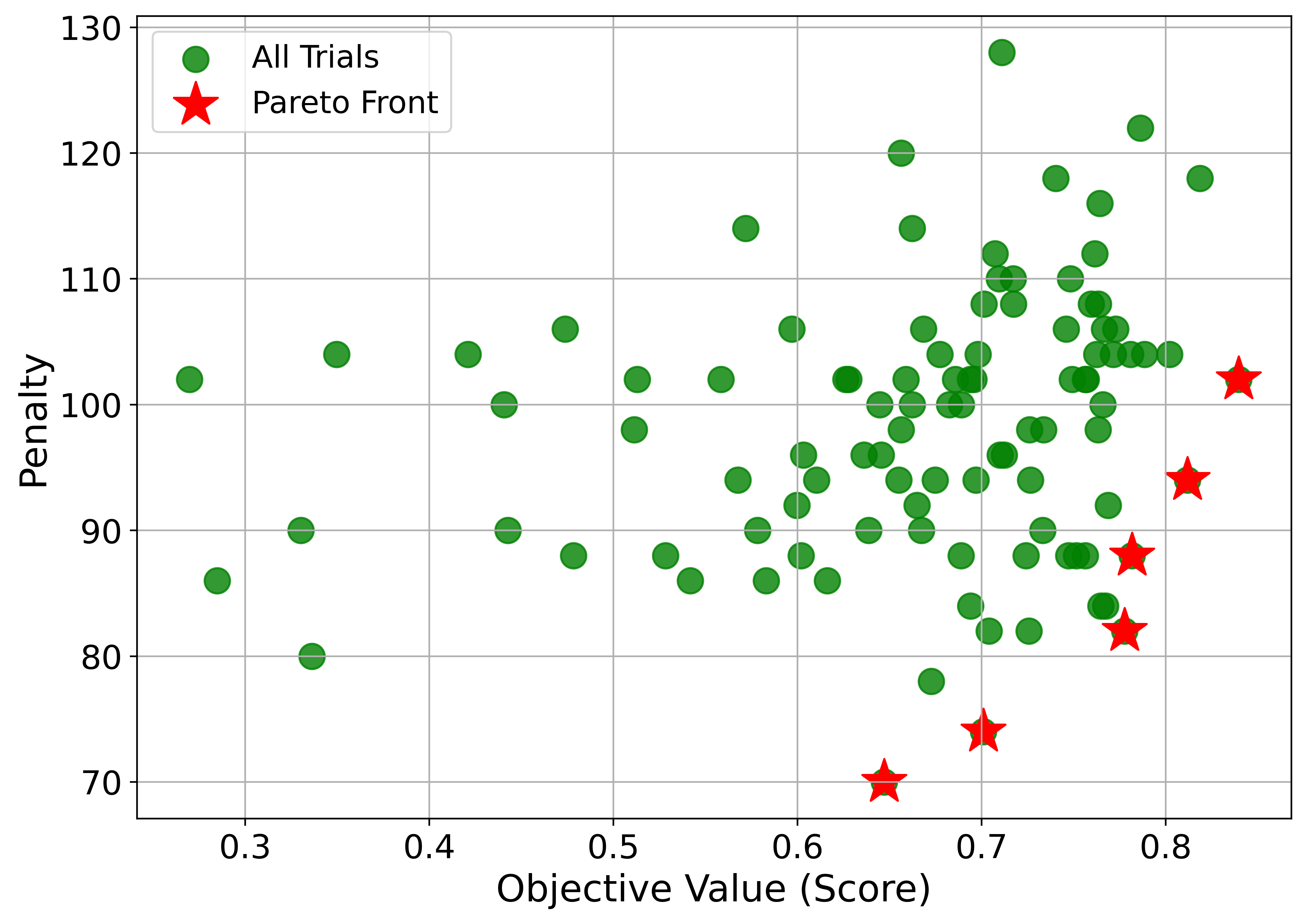}
        \subcaption{}
    \end{minipage}
    \caption{NSGA-II Pareto fronts showing all candidate solutions and the non-dominated Pareto-optimal architectures along the objective–penalty trade-off for a) 4OR configuration, and b) 13OR configuration.}
    \label{fig:NSGA_pareto}
     \vspace{-1mm}
\end{figure}

\begin{table*}[!htp]
\centering
\caption{FPGA Implementation Results of the proposed Architecture design and their comparisons with the existing SOTA works.}
\resizebox{18cm}{!}{%
\setlength{\tabcolsep}{8pt}
\renewcommand{\arraystretch}{3.5}
\begin{tabular}{lllllllllll}
\hline
\Huge Design &
\Huge \# Feature maps &
\Huge FPGA family &
\Huge LUTs &
\Huge Slices &
\Huge \# DSP units &
\Huge \# Multipliers &
\Huge \# Shifters &
\Huge \# Adders &
\Huge Power (mW) &
\Huge Max Freq. (MHz) 


\\ \hline

\Huge \cite{Anbaran} \Huge (FI-16) &
\Huge 4 &
\Huge Virtex 6 &
\Huge 1564 &
\Huge 2248 &
\Huge 156 &
\Huge 72 &
\Huge 992 &
\Huge 198 &
\Huge 892 &
\Huge 100 \\

\Huge \cite{ssimpsnref} \Huge (FI-16) &
\Huge 4 &
\Huge Virtex 7 &
\Huge 3036 &
\Huge 4736 &
\Huge 184 &
\Huge 100 &
\Huge 1014 &
\Huge 218 &
\Huge 984 &
\Huge 100 \\

\Huge \cite{power-eff} \Huge (Int8) &
\Huge 4 &
\Huge Kintex 7 &
\Huge 995 &
\Huge 1002 &
\Huge 0 &
\Huge 0 &
\Huge 609 &
\Huge 238 &
\Huge 403 &
\Huge 100 \\

\Huge This work \Huge (4OR-FI-16) &
\Huge 4 &
\Huge Kintex 7 &
\Huge 8642 &
\Huge 50 &
\Huge 0 &
\Huge 48 &
\Huge 0 &
\Huge 76 &
\Huge 310 &
\Huge 142 \\

\Huge This work\ \Huge (4OR-FI-Graded) &
\Huge 4 &
\Huge Kintex 7 &
\Huge 1308 &
\Huge 50 &
\Huge 0 &
\Huge 48 &
\Huge 0 &
\Huge 76 &
\Huge 115 &
\Huge 142 \\

\Huge This work \Huge (13OR-FI-16) &
\Huge 13 &
\Huge Kintex 7 &
\Huge 21440 &
\Huge 50 &
\Huge 0 &
\Huge 180 &
\Huge 0 &
\Huge 285 &
\Huge 1132 &
\Huge 135 \\

\Huge This work \Huge (13OR-FI-Graded) &
\Huge 13 &
\Huge Kintex 7 &
\Huge 4417 &
\Huge 50 &
\Huge 0 &
\Huge 180 &
\Huge 0 &
\Huge 285 &
\Huge 344 &
\Huge 135 \\
\hline

\end{tabular}%
}

\label{tab:fpga-results}
\vspace{1.5mm}
\begin{minipage}{\textwidth}
\raggedright
\hspace{-0.1cm} Note: \# represents counts, These results are evaluated for 512$\times$512 image size. Power is estimated for a normalized operating frequency of 100MHz. Here, Power refers to the dynamic power of the implementation for a fair comparison.
\end{minipage}
\end{table*}


\begin{table*}[!htp]
\centering
\caption{Optimal GF convolution layer multiplier bit width sequence generated by each NSGA-II that fits best for the PADP objective for 4OR and 13OR configurations.}
\resizebox{18.00cm}{!}{
\renewcommand{\arraystretch}{1.1} 

\begin{tabular}{lllllllll} 

\hline
Implemented Designs       & {\begin{tabular}[c]{@{}c@{}} Multiplier Precisions config \end{tabular}}                            
              & {\begin{tabular}[c]{@{}c@{}} Area \\ ($\mu$m$^2$) \end{tabular}}  
              & {\begin{tabular}[c]{@{}c@{}} Delay \\ (ns) \end{tabular}}     
              & {\begin{tabular}[c]{@{}c@{}} Power \\ (mW) \end{tabular}}  
              & {\begin{tabular}[l]{@{}l@{}} PADP \\ (mW-$\mu$m$^2$ns) \end{tabular}}
            & {\begin{tabular}[l]{@{}l@{}} Accuracy \\ (\%)\end{tabular}} & 
           \multicolumn{2}{c} {\begin{tabular}[l]{@{}l@{}} {{Improvement}}\\(\%)\end{tabular}}\\ 
\cline{8-9}
 &  &  &  &  &  &  & {Area} & {Power} \\ 
\hline
4OR-FI-16        & [16 16 16 16 16 16 16 16 16 16 16 16]  & 106750  & 2.68  & 3.14 &898322.6  & 94.75 & \textemdash & \textemdash \\ 
4OR-FI-graded    & [3  3  3  6  3  4  3  2  5  5  2  2]  & 36193  & 4.06 & 1.33  &  195434.9 &  94.00 & 66.09 & 57.64\\ 
13OR-FI-16       & [16 16 16 16 16 16 16 16 16 16 16 16]  & 245712  & 4.43  &  6.41 & 6966426.6 & 98.60  & \textemdash & \textemdash \\ 
13OR-FI-graded  & [2 4 4 2 3 4 2 5 5 3 4 2]  & 125709  & 3.54  & 4.61&2051495.4 & 98.00  & 48.83 &  28.08 \\ 
\hline
\end{tabular}
}
\label{tab:asic_results}
\vspace{1.5mm}
\begin{minipage}{\textwidth}
\raggedright
\hspace{-0.1cm} Note: The area and power improvement for each configuration with optimized FI-graded precisions is computed against the corresponding baseline designs with FI-16. Here, Power refers to the total power of the circuit.
\end{minipage}
 \vspace{-3mm}
\end{table*}

The evolutionary optimization process is executed to derive Pareto-optimal GF architectures for the base model, which is configured with GF-4OR and GF-13OR, consisting of four and thirteen distinct orientations.
These configurations are passed through a synthesis flow built on a Python framework. Each design is evaluated in terms of area, delay, and power using OpenROAD~\cite{openroad} and the Nandgate 45~nm standard cell library. The hardware parameters used for the experimental setup are listed in the Table~\ref{tab:hardware_parameters}.
This process was carried out over a population size of 100 and 50 generations. For each orientation, the fitness metric was computed using the Euclidean distance across four normalized design objectives: synthesis area, critical path delay, power consumption, and SSIM. The optimization process begins with an initial solution configured using the maximum multiplier width of $9$, and gradually explores a range of varying widths across all multiplier instances during successive iterations. 
Figure \ref{fig:NSGA_pareto} illustrates the Pareto fronts obtained for both the 4OR and 13OR GF configurations. Each scatter plot represents the exploration space generated during evolutionary search, where every point corresponds to a candidate GF architecture evaluated in terms of its objective score (derived from accuracy-driven quality analysis) and penalty (representing a composite measure of area, delay, and power). Pareto-optimal designs in the 4OR case, cluster towards lower penalties while maintaining competitive objective values, indicating that lightweight Gabor configurations still achieves strong feature quality under constrained resources. In contrast, the 13OR case exhibits a broader design distribution, characterized by higher objective values, but also increased penalties, which reflects the additional hardware cost associated with handling more orientations. 
The shape and spread of both plots demonstrate the capability of NSGA-II to efficiently navigate a large architectural search space.
Overall, the Pareto plots validate the effectiveness of the proposed evolutionary framework in identifying optimized GF configurations that deliver high-quality feature extraction while adhering to hardware resource limitations.


To study the efficacy of our approach in approximating the multiplier hardware for the GF, we implemented the GF accelerator architecture for 4OR and 13OR configurations, using both FI-16 and the NSGA-II guided FI-graded precision. The hardware results for the implemented designs were compared with the SOTA work. All hardware characteristics of the implemented design were obtained for both the FPGA synthesis and the ASIC synthesis flow.
For FPGA, performance metrics, logic utilization (LUTs and slices), dynamic power consumption (mW), and maximum operating frequency (MHz) were extracted post place-and-route in Xilinx Vivado 2024.1. To ensure a fair comparison, we re-implemented the SOTA design reported in \cite{power-eff} on the same Kintex-7 (XC7K325T) device and measured it with an identical tool flow. Additional reference data were collected from \cite{Anbaran} and \cite{ssimpsnref}. Table~\ref{tab:fpga-results} lists the FPGA resource usage and arithmetic units in the form of multipliers, shifters, adders, and DSP blocks across all designs. The SOTA approaches in \cite{Anbaran} and \cite{ssimpsnref} utilize FI-16 arithmetic and rely on DSP slices for multiplication, thereby increasing both dynamic power and silicon area. Our design, built entirely with LUTs and slices, eliminates that overhead. Although the solution in \cite{power-eff} occupies slightly less footprint than ours, our design improves in power costs by a remarkable 71.46\% and 14.64\% for 4OR and 13OR configurations, respectively. 

For ASIC synthesis flow, we utilized TCL automation to extract silicon footprint parameters from the Yosys Open-Source Synthesis Suite~\cite{yosys}, and critical path delay and power metrics from the OpenSTA tool~\cite{opensta}, targeting the typical corner conditions for the design using Nangate 45nm OpenCell Library. Power analysis in OpenSTA assumed an input switching activity of 0.25 at a frequency of $100MHz$.  Table~\ref{tab:asic_results} presents the ASIC implementation comparison results for the proposed architecture of FI-16 and FI-Graded precision, as measured by hardware metrics including silicon area, critical path delay, and power cost. Our design for the FI-Graded Precision achieved footprint savings of 66.09\% and 48.83\% and a power benefit of 57.64\% and 28.08\% for 4OR and 13OR configurations, respectively, compared to their corresponding FI-16 implementations. The GF-13OR accelerator, with FI-Graded precision, design offers an excellent classification accuracy of 98\% compared to the GF-4OR design, albeit at the cost of three times more hardware.

\section{Conclusion}~\label{sec:conclusion}
This work presents a method for optimizing the hyperparameters of a GFB using an EfficientNet-B0-based CNN model for medical applications. The optimized GF layer, when integrated with EfficientNet-B0, improves the overall accuracy of the classifier from 96.2\% to 98.6\%. The proposed NSGA-II algorithm-based framework generates quantized fixed-point precision for individual multipliers in the GFB. 
Our hardware-software co-optimization approach achieves efficient hardware metrics improvement with minimal compromise on accuracy. The implementation of GFB with FI-graded coefficients resulted in a silicon footprint savings of 48.83\% and a power improvement of 28.08\% compared to a conventional FI-16 implementation, with a minimal 0.6\% drop in accuracy.
With minimal trade-offs in accuracy, we propose an optimized GF hardware accelerator suitable for efficient feature extraction in GF-based medical image processing applications.
The hardware design files are made freely available at \cite{gaborsite_gaborflow}
for easy adoption and further usage by the researchers and designers. \label{sec:concl}



\bibliographystyle{unsrt}
\bibliography{refs}

\end{document}